\documentclass[%
 aip,
 amsmath,amssymb,
 reprint,%
]{revtex4-1}

\usepackage{graphicx}
\usepackage{dcolumn}
\usepackage{bm}
\usepackage{lineno}
\usepackage[utf8]{inputenc}
\usepackage[T1]{fontenc}
\usepackage{mathptmx}

\begin{document}

\preprint{AIP/123-QED}

\title[The dispersion of spherical droplets in source-sink flows and their relevance to the COVID-19 pandemic]{The dispersion of spherical droplets in source-sink flows and their relevance to the COVID-19 pandemic}

\author{C.P. Cummins}
 \email{c.cummins@hw.ac.uk}
 \affiliation{Maxwell Institute for Mathematical Sciences, Department of Mathematics, Heriot-Watt University, Edinburgh, EH14 4AS, UK}%
\affiliation{Institute for Infrastructure
\& Environment, Heriot-Watt University, Edinburgh, EH14 4AS, UK}%
\author{O.J. Ajayi}%
\affiliation{Maxwell Institute for Mathematical Sciences, Department of Mathematics, Heriot-Watt University, Edinburgh, EH14 4AS, UK}

\author{F.V. Mehendale} 
\affiliation{Centre for Global Health, Usher Institute, College of Medicine and Veterinary Medicine, University of Edinburgh, EH8 9AG, UK}

\author{R. Gabl}%
\affiliation{School of Engineering, University of Edinburgh, Edinburgh, EH9 3FB, UK}%
\author{I.M. Viola}%
\affiliation{School of Engineering, University of Edinburgh, Edinburgh, EH9 3FB, UK}%

\date{\today}

\begin{abstract}
In this paper, we investigate the dynamics of spherical droplets in the presence of a source-sink pair flow field.  The dynamics of the droplets is governed by the Maxey-Riley equation with Basset-Boussinesq history term neglected. We find that, in the absence of gravity, there are two distinct behaviours for the droplets: small droplets cannot go further than a specific distance, which we determine analytically, from the source before getting pulled into the sink. Larger droplets can travel further from the source before getting pulled into the sink by virtue of their larger inertia, and their maximum travelled distance is determined analytically.

We investigate the effects of gravity, and we find that there are three distinct droplet behaviours categorised by their relative sizes: small, intermediate-sized, and large. Counterintuitively, we find that the droplets with minimum horizontal range are neither small nor large, but of intermediate size. Furthermore, we show that in conditions of regular human respiration, these intermediate-sized droplets range in size from a few $\mu$m to a few hundred $\mu$m. The result that such droplets have a very short range could have important implications for the interpretation of existing data on droplet dispersion.

\end{abstract}

\maketitle

\section{\label{sec:intro}Introduction}

The transport of inertial particles in fluid flows occurs in many problems arising in engineering and biology, such as the build-up of microplastics in the ocean\cite{Beron-Vera2019} and respiratory virus transmission through tract droplets\cite{Bourouiba2014,Bourouiba2020,Verma2020}. The Maxey-Riley equation\cite{Maxey1983} describes the motion of a finite-sized spherical particle in an ambient fluid flow. The equation is a representation of Newton's second law, in which the forces acting on the particle include a Stokesian drag force, an added mass force, a gravity force, the force due to the undisturbed flow, and a Basset-Boussinesq history term. The equation takes the form of a second-order, implicit integro-differential equation with a singular kernel, and with a forcing term that is singular at the starting time\cite{Farazmand2015}. The equation has been applied to model the dynamics of aerosol comprising particles of various density ratios\cite{Maxey1987}, feeding mechanism of jellyfish\cite{Costello1994,Peng2009}, and the dynamics of inertial particles in vortical flows\cite{Druzhinin1994,Raju1997,Ravichandran2015}. The equation has also been applied to droplet-laden flows with a phase change at sub-Kolmogorov scales\cite{Ravichandran2020}.

The Basset-Boussinesq term accounts for the drag due to the production of vorticity as the particle is accelerated from rest. It is difficult to include this term numerically, and is often omitted on the assumption that particles move in a quasistatic manner\cite{Michaelides1997}. This assumption breaks down in bubbly and slurry flows, where the Basset-Boussinesq term accounts for a quarter of the forces on the particle\cite{Michaelides1997} when density ratio $R={2\rho_{\text{f}}'}/\left(\rho_{\text{f}}'+2\rho_{\text{p}}'\right)$ is greater than $2/3$, where $\rho_{\text{f}}'$ is the fluid density and $\rho_{\text{p}}'$ is the particle density. Recent advances\cite{Prasath2019} have shown that the full Maxey-Riley equation can be represented as a forced, time-dependent Robin boundary condition of the 1-D diffusion equation. Here, the authors found that a particle settling under gravity relaxes to its terminal velocity according to $t^{-1/2}$; however, if the Basset-Boussinesq term is neglected, it relaxes exponentially quickly\cite{Langlois2015}.

In this paper, we examine the transport of inertial particles in source-sink flows\cite{Batche2000}. Such a flow could represent the trajectories of water droplets emitted from coughing, sneezing\cite{Bourouiba2014,Bourouiba2020,Verma2020}, or breathing and in the presence of extraction, such as an air-conditioning unit or air current\cite{Dbouk2020}. Our simplified mathematical model yields to analytic treatment in certain limits of large and small droplets. This enables us to provide important physical insight into this complex problem, but we remark that effects such as drag non-linearity\cite{Rosa2016} and turbulent dispersion\cite{Busco2020} are not taken into account. 
Since the dynamics of settling droplets is significantly affected by their size, it is important to understand the impact that the emitted droplet size has on the destination of such a droplet in a source-sink flow. In particular, since droplets are vectors for infectious diseases such as COVID-19, it is imperative that we understand the droplet dynamics in such flows to mitigate the spread of the disease.

The paper is organised as follows: in Section~\ref{sec:maths-model}, the mathematical model is presented and non-dimensionalised. The results are presented in Section~\ref{sec:results}, for small (\ref{sec:verysmall}) and intermediate-sized (\ref{sec:medsized}) droplets in the absence of gravity. Gravitational effects are considered for small droplets in \ref{sec:gravity} and for intermediate-sized droplets in \ref{sec:gravitymedsized}. In Section~\ref{sec:physicalextractor}, we present applications of our results for human breathing without (\ref{sec:drop_disp}) and with (\ref{sec:extractor}) the inclusion of extraction. Finally, we discuss our findings in Section~\ref{sec:conclusion}.

\section{\label{sec:maths-model}Mathematical model}

Consider a source producing air of density $\rho_{\text{air}}'$ and viscosity $\nu_{\text{air}}'$, with volume flux of $Q_{1}'$, containing spherical liquid droplets of density $\rho_\text{drop}'$, which are emitted with a characteristic velocity $U'$. Let us represent the 3D velocity field $\mathbf{u'_{\text{source}}}(\mathbf x')$ at a position $\mathbf x'$ of the emitted air as a point source of strength $Q_{1}'$, centred at the origin in Cartesian coordinates\cite{Batche2000}:
\begin{equation}
\mathbf{u_{\text{source}}'}(\mathbf x')= \frac{Q_{1}'\mathbf x'}{4\pi |\mathbf x'|^{3}}.
\end{equation}
We include an extraction unit as a point sink of strength $Q_{2}'$ located at a position $\mathbf x_0'$ as follows:
\begin{equation}
\mathbf{u_{\text{sink}}'}(\mathbf x')=- \frac{Q_{2}'(\mathbf x'-\mathbf x_0')}{4\pi |\mathbf x'-\mathbf x_0'|^{3}}.
\end{equation}
The resulting airflow is given by the linear superposition of these two flows:
\begin{equation}
\mathbf{u'}(\mathbf x')=\frac{Q_{1}'\mathbf x'}{4\pi |\mathbf x'|^{3}}- \frac{Q_{2}'(\mathbf x'-\mathbf x_0')}{4\pi |\mathbf x'-\mathbf x_0'|^{3}}.
\label{eq:dim-Vel-Field}
\end{equation}
The natural timescale of the problem emerges as $T'=|\mathbf x_0'|/U'$. We non-dimensionalise \eqref{eq:dim-Vel-Field} according to 
\begin{equation}
\mathbf{x}=\mathbf{x'}/|\mathbf x_0'|\quad \mathbf{u}=\mathbf{u'}/U',
\label{eq:u-X}
\end{equation}
which gives the nondimensionalised expression for the airflow velocity
\begin{equation}
\mathbf{u}(\mathbf x)=\Lambda \left(\frac{\mathbf x}{|\mathbf x|^{3}}- \gamma\frac{(\mathbf x-\mathbf x_0)}{|\mathbf x-\mathbf x_0|^{3}}\right),
\label{eq:non-dim-Vel-Field}
\end{equation}
with $\Lambda= {Q_{1}'}/{4\pi  U' |\mathbf x_0'|^{2}}$,  $\gamma={Q_{2}'}/{Q_{1}'}$, and  $\mathbf x_0=\mathbf x'_0/|\mathbf x'_0|$.

The velocity of the droplet embedded in this background airflow obeys the Maxey-Riley equation\cite{Maxey1983}
\begin{multline}
 \dot{\mathbf{v}}(t)-\frac{3}{2}R\frac{\text{D}\mathbf{u}}{\text{D}t}\Big|_{\textbf{X}(t)}=\left(1-\frac{3}{2}R\right)\mathbf{g}-A\left(\mathbf{v}(t)-\mathbf{u}(\textbf{X}(t),t)\right)\\
 -\sqrt{\frac{9}{2\pi}}\frac{R}{\sqrt{St}}\left[\int_{0}^{t}\frac{\dot{\mathbf{v}}(s)-\dot{\mathbf{u}}(\textbf{X}(s),s)}{\sqrt{t-s}}\text{d}s+\right. \\
\left. \frac{\mathbf{v}(0)-\mathbf{u}(\textbf{X}(0),0)}{\sqrt{t}}\right].
\end{multline}
where $\mathbf X(t)$ is the position of the droplet at time $t$, $\textbf v(t) = \dot{\textbf{X}}(t)$ is its velocity, the dot indicates time derivative, and
\begin{multline}
R=\frac{2\rho_{\text{air}}'}{\rho_{\text{air}}'+2\rho_{\text{drop}}'}, \quad  A=\frac{R}{St},\\ St=\frac{2}{9}\left( \frac{a'}{|\mathbf x_0'|}\right)^2 Re,\quad \mathbf g =\frac{|\mathbf x_0'|\mathbf g'}{U'^2},
\label{eq:MR-parameters}
\end{multline}
with $a'$ the droplet radius, $\mathbf g'$ the acceleration due to gravity vector, $Re=U'|\mathbf x_0'|/\nu_{\text{air}}' $  is the Reynolds number, and $St$ is the particle Stokes number. Note here that the Fax\'en correction terms\cite{Maxey1983} have not been omitted: they are identically zero since $\Delta \mathbf u=\mathbf 0.$

The approximate ratio of the Basset history drag to Stokes drag is $O(St^{1/2})$, which, for the range of $St$ we are interested in, is generally much less than one. In the remainder of the paper, we neglect the Basset history term since we anticipate its magnitude is negligible compared to the Stokes drag term for the parameters of interest to us, and the resulting equations are
\begin{equation}
 \dot{\mathbf{v}}(t)-\frac{3}{2}R\frac{\text{D}\mathbf{u}}{\text{D}t}\Big|_{\textbf{X}(t)}=\left(1-\frac{3}{2}R\right)\mathbf{g}-A\left(\mathbf{v}(t)-\mathbf{u}({\textbf{X}(t)},t)\right),
\label{eq:MR}
\end{equation}
subject to the initial conditions $\mathbf v(0)=\mathbf u(\mathbf X(0),0)$ where $\mathbf X(0)$ lie on a circle surrounding the origin of radius $|\mathbf X(0)|$. In \eqref{eq:non-dim-Vel-Field}, taking the limit
\begin{equation}
\lim_{\mathbf X \to \mathbf 0}\mathbf{u}(\mathbf X)\simeq  \frac{\Lambda\mathbf X}{|\mathbf X|^{3}};
\label{eq:MRX0}
\end{equation}
hence, we can ensure that the non-dimensional initial velocity has unit magnitude by requiring $|\mathbf X(0)|=\sqrt{\Lambda}$.

\subsection{\label{sec:comp} Computational considerations}

The resulting equations \eqref{eq:MR} are a set of three coupled second-order non-linear ordinary differential equations (ODEs) for the position vector $\mathbf X(t)$. The algebra involved in computing the material derivative in \eqref{eq:MR} is straightforward, but cumbersome, and it is omitted here. This set of equations does not admit analytical solutions in general, so it must be solved numerically. 

We solved the equations by expressing them as a system of six first-order ODEs using the MATLAB\textsuperscript{\textregistered}  ode15s algorithm, a variable-step, variable-order solver based on the numerical differentiation formulas\cite{Shampine1997}. This was performed on a laptop equipped with an Intel(R) Core(TM) i9-9980HK CPU (2.40GHz) and 32GB of RAM; and each trajectory took on average 0.015 seconds to compute. In each of our plots, we show the trajectories emanating from 30 evenly-spaced points on a circle centred at the origin (i.e., the source), giving a total simulation time of approximately 0.45 seconds. The number of trajectories was selected by balancing the requirements on the detail on individual trajectories and the global behaviour of the droplets. Such short simulation times allows us to identify the most important combinations out of a wide variation of variables in a computational time that is several orders of magnitude faster than models employing computational fluid dynamics\cite{Xi2016}.  


\section{\label{sec:results}The results}

\subsection{\label{sec:verysmall} Small droplets in the absence of gravity}

In the absence of gravity, \eqref{eq:MR} read (dropping the explicit time dependence)
\begin{equation}
 \dot{\mathbf{v}} -\frac{3R}{2} \left[\mathbf u \cdot \nabla \mathbf u \right]\Big|_{\textbf{X}}=-\frac{R}{St}\left( \mathbf v-\mathbf u \Big|_{\textbf{X}(t)}\right).
\label{eq:MR-small-st}
\end{equation}
In \eqref{eq:MR-small-st}, for small droplets ($St\ll R$) emitted from the source, the balance is between the first term on the left-hand side and the right-hand side, so that the velocity rapidly adjusts to the background flow $ \mathbf v \approx  \mathbf u \Big|_{\textbf{X}(t)}$.

We are interested in whether droplets move away from or towards the sink. To this end, we look for trajectories for which $\mathbf v>0$:
\begin{equation}
\mathbf v=\frac{\text{d} \mathbf X}{\text{d} t}>\mathbf 0 \iff \frac{\mathbf X}{|\mathbf X|^{3}}> \gamma\frac{(\mathbf X-\mathbf x_0)}{|\mathbf X-\mathbf x_0|^{3}}.
\end{equation}
If we take $\mathbf x_0=\left[1,0,0\right]$, then the trajectory that emerges from the source and travels in the direction of the negative $x$-axis is the one that gets the greatest distance away from the sink. Hence, let us consider this inequality in the first component, and along the line $y=0$, $z=0$:
\begin{equation}
\frac{\text{d} X(t)}{\text{d} t}> 0 \iff \frac{X}{|X|^{3}}> \gamma\frac{(X-1)}{|X-1|^{3}}.
\end{equation}
We are interested in where the flow field changes direction, since this indicates the maximum distance droplets emitted at the source can travel before moving towards the sink. To this end, let us choose a point $x=-\lambda$ along $y=0$ and $z=0$; then this inequality tells us that
\begin{equation}
\frac{\text{d} X(t)}{\text{d} t}> 0 \iff \gamma>\left( 1+\frac{1}{\lambda}\right)^2.
\end{equation}
This inequality can hold only if $\gamma>1$. This makes sense, since flow is directed towards the sink only if the sink is stronger than the source. 

\begin{figure*}
\includegraphics[width=0.85\textwidth]{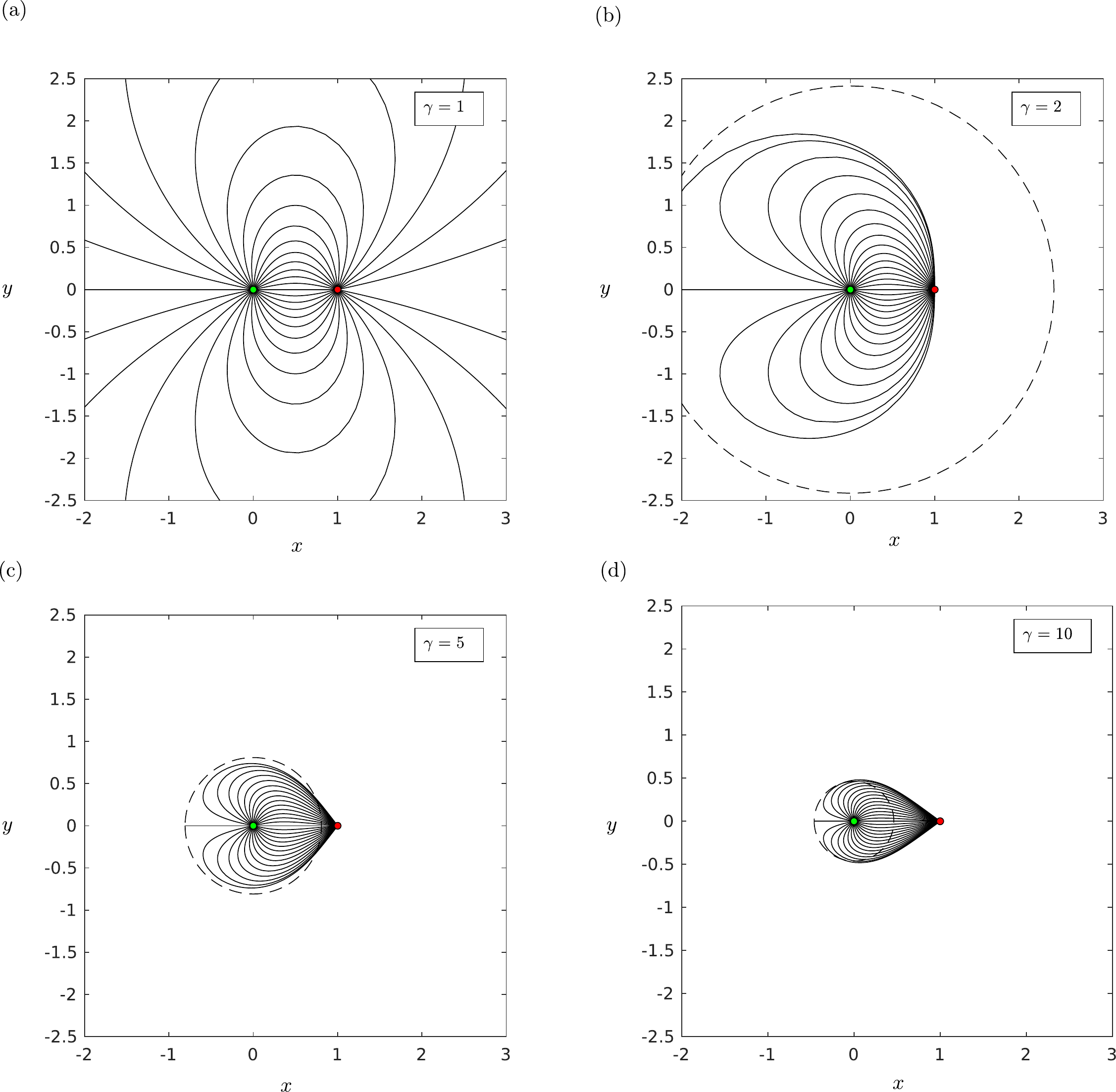}
\caption{\label{fig:gamma-study} The trajectories $\mathbf X(t)$ in the $xy$-plane of small droplets $St\ll R$ with a background source-sink pair of various strengths: (a) $\gamma=1$,  (b) $\gamma=2$,  (c) $\gamma=5$, and  (d) $\gamma=10$. In these plots, $R=0.001$, $\Lambda=0.0001$, $|\mathbf{g}|=0$. The trajectories do not change for changing $R$. The dashed circle indicates the predicted maximal distance that a droplet can travel in this regime, calculated using the inequality \eqref{eq:lambda-gamma}. The source is indicated by a green filled circle and the sink is indicated by a red filled circle.}
\end{figure*}

Figure~\ref{fig:gamma-study} shows the trajectories for small droplets ($St\ll R$) in the presence of a source-sink pair: the source is located at the origin (green disk) and the sink is located at $x=1$ along the $x$-axis (red disk). For $\gamma=1$ (Figure~\ref{fig:gamma-study}a), we have equal strength and droplets can take large excursions from the source before returning to the sink. As $\gamma$ increases, the trajectories emanating from the source occupy an increasingly compacted region (Figure~\ref{fig:gamma-study}b-d). We can use this inequality above to define a region 
\begin{equation}
|\lambda|<\frac{\sqrt{\gamma }+1}{\gamma -1},
\label{eq:lambda-gamma}
\end{equation}
such that small droplets do not get further than a distance $|\lambda|$ before travelling towards the sink. The circle with radius $|\lambda|$ is shown in Figure~\ref{fig:gamma-study} (dashed curve). Observe that, as one gets increasingly close to the source ($\lambda\to 0$), the inequality tends to 
\begin{equation}
\frac{\text{d} X(t)}{\text{d} t}> 0 \iff \gamma>\frac{1}{\lambda^2},
\end{equation}
meaning that, in order to maintain trajectories moving away from a given test point, the sink strength needs to increase quadratically with distance of the test point to the source.

\subsection{\label{sec:medsized} Intermediate-sized droplets in the absence of gravity}

For $St =O(R)$ and $St \gg R$, the droplet is slowed down exponentially according to
\begin{equation}
\mathbf v(t)\approx\mathbf v(0) \exp{\left[-(R/St)t\right]},
\end{equation}
which represents a balance between inertia and drag forces. Provided $\gamma>1$, and in the absence of gravity, in the long-term, the droplet will always migrate towards the sink. However, in the case of intermediate-sized droplets, the maximum distance travelled by the droplet before it moves towards the sink is given by $|\mathbf v(0)|/(R/St)$. Since the initial velocity of the droplet is chosen to be the same as the surrounding fluid, then we can write the maximum distance as $|\mathbf u(\mathbf X(0),0)|/(R/St)$. As explained above (see \eqref{eq:MRX0}), in our non-dimensionalisation, our characteristic velocity $U'$ was chosen to be that of the outlet. Hence, in this non-dimensionalisation, $|\mathbf u(\mathbf X(0),0)|=1$.  

\begin{figure}
\includegraphics[width=0.4\textwidth]{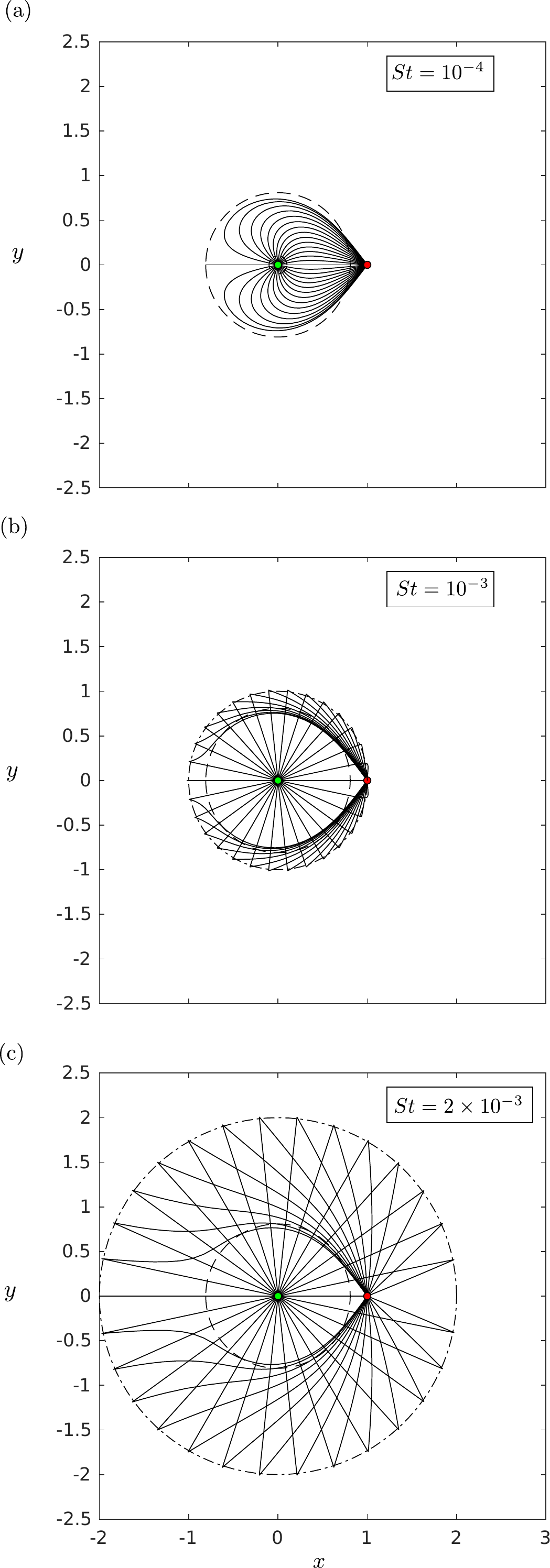}
\caption{\label{fig:St-study-exponential} The trajectories $\mathbf X(t)$ in the $xy$-plane of droplets with a background source-sink pair with strength ratio $\gamma=5$ for various values of $St$: (a) $St=10^{-4}$,  (b) $St=10^{-3}$, and  (c) $St=2\times10^{-3}$. In these plots, $R=0.001$, $\Lambda=0.0001$, $|\mathbf{g}|=0$. The dashed circle indicates the predicted maximal distance that a droplet can travel in this regime, calculated using the inequality \eqref{eq:lambda-gamma}. The dash-dotted circle indicates the maximal distance predicted by inertia-drag balance, giving radius equal to $St/R$. The source is indicated by a green filled circle.}
\end{figure}

Figure~\ref{fig:St-study-exponential} shows the trajectories of intermediate-sized droplets for $\gamma=5$ in the absence of gravity. The striking feature of the plot is the shift from a regime where the maximal extent of the trajectories as predicted by \eqref{eq:lambda-gamma} is no longer valid and must be replaced with a circle of radius $St/R$. In Figure~\ref{fig:St-study-exponential}a, $St/R=0.1$, so that the droplets are slowed down rapidly before following the fluid flow. In Figure~\ref{fig:St-study-exponential}b, $St/R=1$, meaning that the droplets are slowed down over the area covered by the unit circle, before being brought to the sink as ideal tracers. Finally, in Figure~\ref{fig:St-study-exponential}c, $St/R=2$ so that the droplets travel a non-dimensional distance of 2 before being slowed down enough to be pulled into the sink. 

Hence, we find that, in the absence of gravity, we can have two very different behaviours depending on whether we have small droplets $St\ll R$ or intermediate-sized droplets $St\geq R$. Small droplets cannot get further than a distance $(\sqrt{\gamma }+1)/(\gamma -1)$ from source before travelling towards the sink, but intermediate-sized droplets are not restricted by this, and can travel further than this provided $St/R>(\sqrt{\gamma }+1)/(\gamma -1)$.

\subsection{\label{sec:gravity} The effect of gravity on small droplets}

As droplets move from the source to the sink, gravity attempts to pull them vertically downwards. Over the timescale of the problem: i.e., the average time it takes a droplet to travel from source to sink, gravity may or may not have an appreciable effect. Intuitively, one would imagine that smaller droplets are influenced more by the airflow than gravity: for stronger sinks, the effect of gravity is comparatively less. Also, intuitively, one would expect that this holds true provided that the source and sink are not too far away. The gravitational vector is non-dimensionalised according to $U'^2/|\mathbf x_0'|$ as shown in \eqref{eq:MR-parameters} so it depends on the the initial speed, and the distance between the source and sink. 

\begin{figure*}
\includegraphics[width=0.85\textwidth]{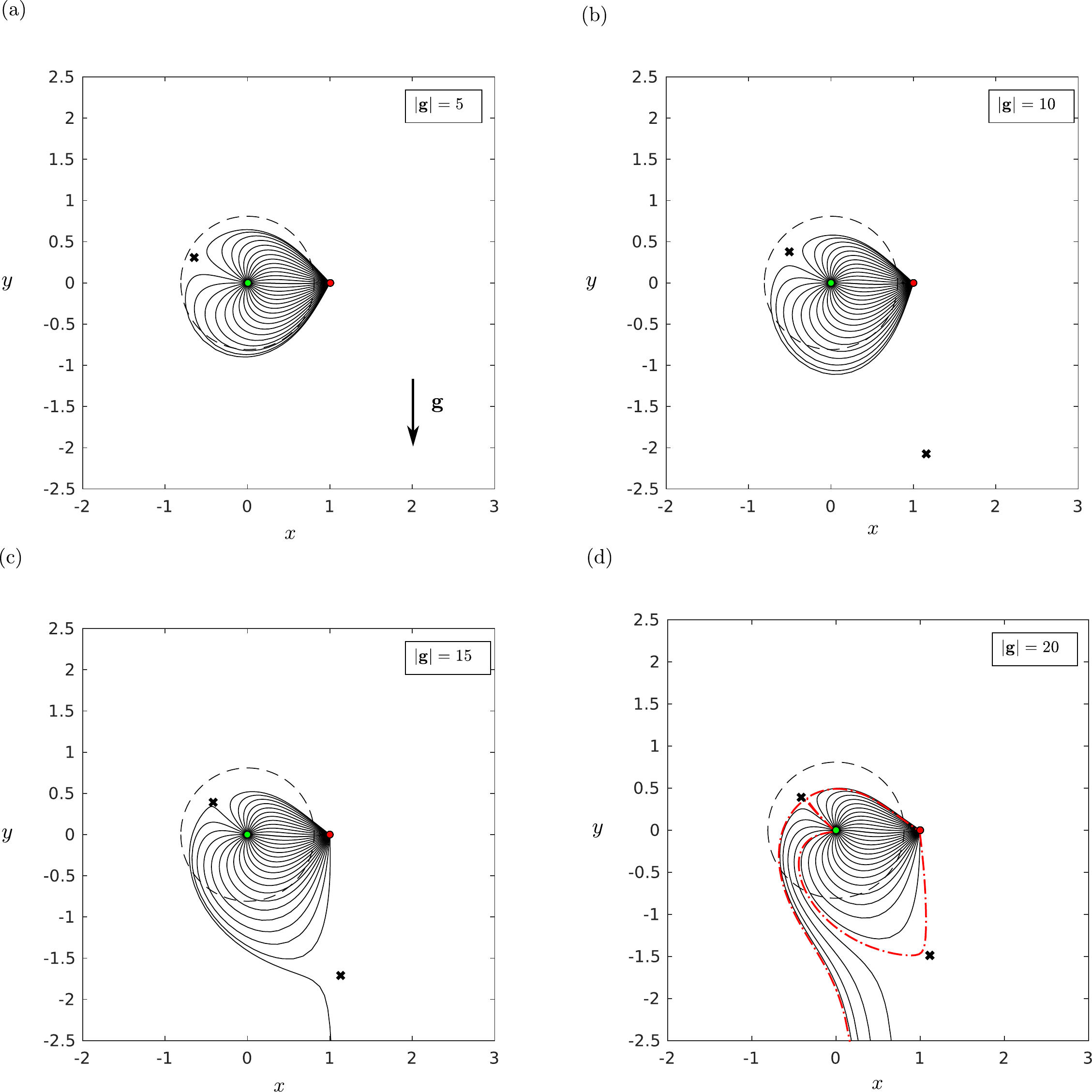}
\caption{\label{fig:gravity-study} The trajectories $\mathbf X(t)$ in the $xy$-plane of small droplets $St\ll R$ with a background source-sink pair with strength ratio $\gamma=5$ and for various strengths of the gravity parameter: (a) $|\mathbf g|=5$, (b) $|\mathbf g|=10$, (c) $|\mathbf g|=15$ and (d) $|\mathbf g|=20$. In these plots, $R=0.001$, $\Lambda=0.0001$. The trajectories will be different for different choices of $R$. The dashed circle indicates the predicted maximal distance that a fluid parcel can travel when ejected from the source. The black crosses indicate the position of saddle fixed points. The source is indicated by a green filled circle and the sink is indicated by a red filled circle.}
\end{figure*}

For $St\ll R<2/3$, and in the absence of gravity, there are three fixed points: the source, sink and a saddle point located at $x=-|\lambda|$ along the $x$-axis (Figure~\ref{fig:gamma-study}). When gravitational effects are included, the fixed point at $x=-|\lambda|$ moves clockwise along about the origin as the effect of gravity is increased (see Figure~\ref{fig:gravity-study}a). A fourth fixed point (saddle) is created far from the source-sink pair, which gradually moves towards the sink (Figure~\ref{fig:gravity-study}b,c) as the effect of gravity is increased. In Figure.~\ref{fig:gravity-study}d, the separatrices (indicated as the red dash-dotted curves) show that there is a wedge of trajectories that escape the pull of the sink. As might be expected, these trajectories are those that point directly away from the sink. Our results show that even for small droplets, gravity can be important if either the sink is far away or if the ejection speed is too low. 

\subsection{\label{sec:gravitymedsized} The effect of gravity on intermediate-sized droplets}

\begin{figure*}
\includegraphics[width=0.85\textwidth]{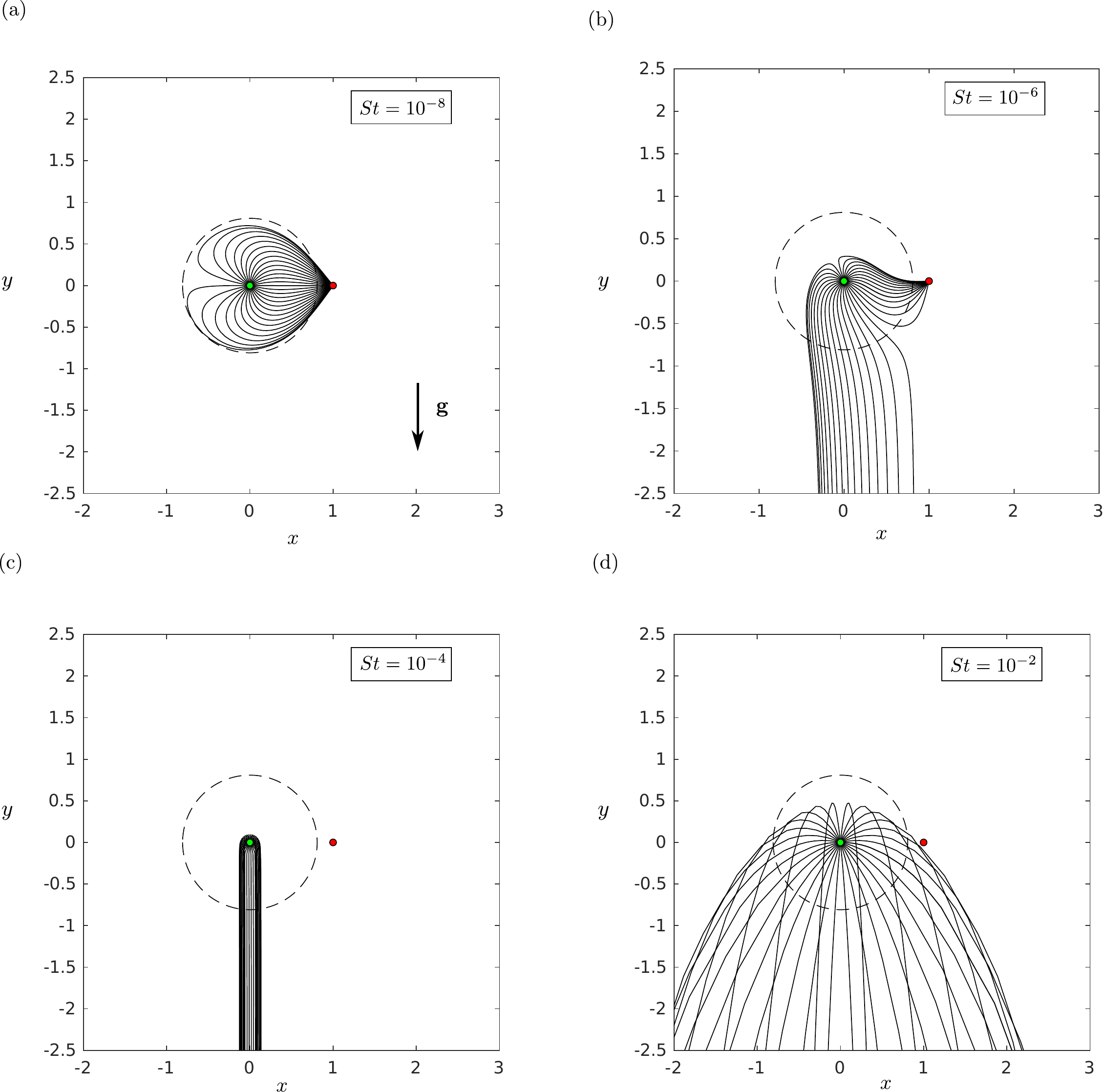}
\caption{\label{fig:gravity-study-St} The trajectories $\mathbf X(t)$ in the $xy$-plane with a background source-sink pair with strength ratio $\gamma=5$ and for various Stokes numbers $St$: (a) $St=10^{-8}$, (b) $St=10^{-6}$, (c) $St=10^{-4}$, and (d) $St=10^{-2}$.   In these plots, $R=0.001$, $\Lambda=0.0001$, and $|\mathbf g |=1$. The trajectories will be different for different choices of $R$. The dashed circle indicates the predicted maximal distance that a fluid parcel can travel when ejected from the source. The source is indicated by a green filled circle and the sink is indicated by a red filled circle.}
\end{figure*}

Small droplets  are deflected by gravity, but generally feel the pull of the sink. Whether or not they are pulled in is determined by the interaction of gravity, the angle of their trajectory, and $\gamma$. As the droplets become larger, gravitational effects dominate, and the sink becomes ineffective. In Figure~\ref{fig:gravity-study-St}, we show how the droplet trajectories behave as $St$ is increased. Figure~\ref{fig:gravity-study-St}a, show the familiar situation where the droplets are so small that gravity does not appreciably affect their trajectory over the characteristic lengthscale.

As $St$ is increased, Figure~\ref{fig:gravity-study-St}b shows that there are a range of trajectories with ejection angles $\alpha$ (defined with respect to the positive sense of the $x$-axis) around the source that are deflected downwards away from the sink. This is consistent with previous sections. However, at a critical $St\approx 2.5\times 10^{-6}$, each trajectory is deflected downwards by gravity (Figure~\ref{fig:gravity-study-St}c). In this case, the maximum horizontal distance travelled by the droplets is very small. Interestingly, this trend is not monotonic. Further increasing $St$, the trajectories adopt a ballistic trajectory (Figure~\ref{fig:gravity-study-St}d). Such droplets can move in very close proximity to the sink, but are not pulled into it (Figure~\ref{fig:gravity-study-St}d).

\section{\label{sec:physicalextractor} Examples of Application}

\subsection{Background on Respiratory Virus Transmission}

One of the possible applications of this paper, is to underpin more sophisticated analytical or numerical models to study the transmission of respiratory viruses. In medical applications, it is common practice to categorise the emitted fluid particles as larger droplets from 5 $\mu$m to 1 mm in diameter that have a ballistic trajectory, and aerosol that remains airborne\cite{Mittal2020}. Droplets smaller than 5 $\mu$m and desiccated droplet nuclei are known as aerosol, which can remain airborne for several hours\cite{Nicas2005,Tellier2006,VanDoremalen2020}. Respiratory viruses are transmitted from virus-laden fluid particles to the recipient through (1) aerosol inhalation; or (2) droplet deposition on the recipient’s mouth, nose or conjunctiva; or (3) droplet deposition on a surface and successive transmission through physical contact\cite{Jones2015}. The SARS-CoV-2 virus, for example, has a diameter of 70–90 nm\cite{Kim2020} and it is carried by droplets and aerosol\cite{VanDoremalen2020,Liu2020}.

The model proposed in this paper can provide new insights on the aerosol transmission, i.e. through those particles whose Stokes number is not sufficiently large to have a ballistic trajectory.  The relative importance of aerosol (1) and droplet (2 and 3) virus transmission is not always known, and it is yet to be established for the SARS-CoV-2\cite{Leung2020}. Counterintuitively, it has been argued that aerosol could be more dangerous than larger droplets\cite{Zhang2020}. Smaller droplets ($\leq$5 $\mu$m) suspended in aerosol might carry a higher concentration of virus than larger droplets (>5 $\mu$m)\cite{Milton2013,Lindsley2015,Leung2020}. The largest droplets are less likely to penetrate deeply in the respiratory system and might be deactivated by the effective first structural and defence barrier of the mucosa\cite{Fokkens2000}. Conversely, aerosolised virus half-life exceeds one hour\cite{VanDoremalen2020} and can be transported airborne through inhalation deep into the lungs\cite{Atkinson2009a,Zayas2012,Yan2018,Lindsley2010}, avoiding the defences of the upper respiratory system. Furthermore, aerosol inoculation has been shown to cause more severe symptoms than droplets administered by intranasal inoculation and the dose of influenza required for inoculation by the aerosol route is 2-3 orders of magnitude lower than the dose required by intranasal inoculation\cite{Lindsley2010,Lindsley2015,Bourouiba2014}. 

To apply our model to aerosol dispersion, we consider the particles ejected by a person talking. A person ejects about tens of fluid particles per second with diameters between\cite{Asadi2019} 0.1 $\mu$m to 1 mm and with a speed of the order\cite{Tang2013} of 1 m s$^{-1}$. Because this is the most frequent source of aerosol, this accounts for most of the aerosol inhaled by other people\cite{Fiegel2006,Atkinson2008}. Coughing leads to the ejection of 100-1000 fluid particles per second with a speed around 10 m s$^{-1}$, while sneezing generates 1000-10,000 fluid particles per second with a speed of up to\cite{Han2013} 20 m s$^{-1}$. The values presented in this paragraph should be taken as indicative because there is a significant variability between different experimental studies\cite{Duguid1945,Duguid1946,Loudon1967,Papineni1997,Morawska2006,Morawska2009,Yang2007,Chao2009,Scharfman2016,Stadnytskyi2020a,Xie2009,Johnson2011,Zayas2012,Lindsley2010,Lindsley2015,Bourouiba2014}.

Some of the physics that is not considered in this work, is the particle-particle interaction and evaporation. In fact, fluid particles are ejected through a jet that transports particles in the range of 2 $\mu$m – 150 $\mu$m \cite{Wells1934, Duguid1946, Xie2007}, i.e. the aerosol, while the largest droplets have a ballistic trajectory independent of the surrounding flow \cite{Wei2015,Xie2007,Bourouiba2014}. The jet can be either laminar or turbulent when breathing and speaking, while coughing and sneezing always results in a turbulent jet with a diameter-based Reynolds number higher\cite{Bourouiba2014} than $10^4$. Once ejected, the air jet extends along a straight trajectory; its diameter increases linearly with the travelled distance, while the mean velocity linearly decreases, and the turbulent statistics remain constant (i.e. the jet is self similar\cite{Morton1956}). Once the largest particles with a ballistic trajectory have left the air jet, the jet bends upwards due to the buoyancy force caused by the temperature and thus density difference\cite{Bourouiba2014}. Smaller size particles ($\leq$100 $\mu$m) are transported by the jet while they evaporate. Once a droplet exits the jet, it falls at its settling speed. For a particle with a diameter of 50 $\mu$m and 10 $\mu$m, the settling speed is less than 0.06 m s$^{-1}$ and 0.03 m s$^{-1}$, respectively. The smallest of these two droplets is likely to land in the form of a desiccated nucleus. In fact, while a droplet with a diameter of 50 $\mu$m
evaporates in about 6 s, a 10 $\mu$m droplet  evaporates in less than\cite{Holterman2003,Bourouiba2014} 0.1 s, although their survivability also depends on the ambient temperature and relative humidity\cite{Busco2020}. Once these droplets leave the jet, they can still be transported by ambient air currents, which have speeds typically in excess\cite{Melikov2012} of 0.01 m s$^{-1}$. These currents are modelled by the sink-source flow field discussed in this paper. 

A key issue that is discussed in this study is the extent to which the cloud of droplets and aerosol are displaced into the neighbouring environment, as this is associated with virus transmission risk. Previous studies estimated that the overall horizontal range of the droplets generated while breathing and coughing before they land on the ground is around 1-2 m\cite{Wells1934,Xie2007,Wei2015}. These studies led to the CDC\cite{CDC2020} and WHO\cite{WHO2020} social distancing guidelines. Nonetheless, the complex physics involved, which includes knowledge of the particle size distribution, their speed of evaporation, the viral charge of droplets of different size, the diffusivity of the virus-laden particles, etc., makes it difficult to assess what is the effective dispersion of virus-laden fluid particles into the environment once ejected. It was found that the largest droplets generated by sneezing can reach a distance as high as 8 m\cite{Bourouiba2014,Bourouiba2020,Xie2007}, while aerosol dispersion is highly dependent on the temperature, humidity and air currents. For these reasons, this paper does not aim to provide definitive measures for the aerosol displacements but contributes to building a body of evidence around this complex question.

\subsection{Predicted Droplet Dispersion} \label{sec:drop_disp}

Currently, there is a large amount of disagreement in the reported spectra of droplet sizes in respiratory events\cite{Bourouiba2014}. The analysis is complicated by various factors including the evaporation of the droplets as they travel from the source, which in turn, is influenced by ambient humidity and temperature. Recent mathematical modelling of droplet emission during talking have categorised droplets into one of three groups\cite{Chen2020}: small ($<75\mu$m), intermediate ($75$-$400\mu$m) and large ($>400\mu$m). Small droplets approximately follow the air and can travel a great distance by weakly feeling the effects of gravity. Large droplets can also travel a large distance due to their inertia. However, the intermediate-sized droplets feel strongly both gravity and drag, and their trajectory is a complex interaction of these effects. Similar trends were observed in computational fluid dynamics simulations of previous authors\cite{Zhu2006}.

In this section, we examine the problem from a much simplified perspective: we ignore evaporation entirely. We model the situation as a point source emitting droplets of various sizes in the presence of gravitational forces, and compute the maximum horizontal distance travelled by these droplets. In this case, $Q_2'=0$ L min$^{-1}$, and other quantities such as jet speed, direction and spread are taken from recent experimental studies of the authors\cite{Viola2020}: these quantities are summarised in Table \ref{tab:dispersion}.

\begin{table}[]
    \centering
\begin{tabular}{llll}
\hline
Quantity    & Description & Value & Units \\
\hline
$U'$      & Jet velocity (quiet)$\dagger$   & 0.55 &   m s$^{-1}$  \\
   & Jet velocity (heavy)$\dagger$    & 4.97 &   m s$^{-1}$  \\
   $\alpha$ & Jet angle (direction)$\dagger$      &-5.8     & $^\circ$ \\
      $\beta$ & Jet angle (spread)$\dagger$     &29.2    & $^\circ$ \\
$\rho_{\text{air}}'$          & Density of air  & 1.149      & $\text{kg} \,\text{m}^{-3}$\\
$\rho_{\text{drop}}'$     & Density of droplet     & 1000     & $\text{kg} \,\text{m}^{-3}$ \\
$\nu_{\text{air}}'$      & Viscosity of air     & $16.36\times 10^{-6}$     & $\text{m}^2 \,\text{s}^{-1}$\\
$Q_1'$ & Volume influx (quiet)$\dagger$      &23.8    & L min$^{-1}$ \\
& Volume influx (heavy)$\dagger$      &133    & L min$^{-1}$ \\
$Q_2'$ & Volume outflux      &0     & L min$^{-1}$ \\
$|\mathbf x_0'|$ & Characteristic length$\ddagger$     &0.5     & m \\
\hline
\end{tabular}
\caption{Physical quantities for dispersion of droplets. $\dagger$ parameters taken from previous experimental study\cite{Viola2020,Viola2020DS01} and $\ddagger$ taken from wind tunnel experiments\cite{Johnson2011}.} \label{tab:dispersion}
\end{table}

\begin{figure*}
\includegraphics[width=0.95\textwidth]{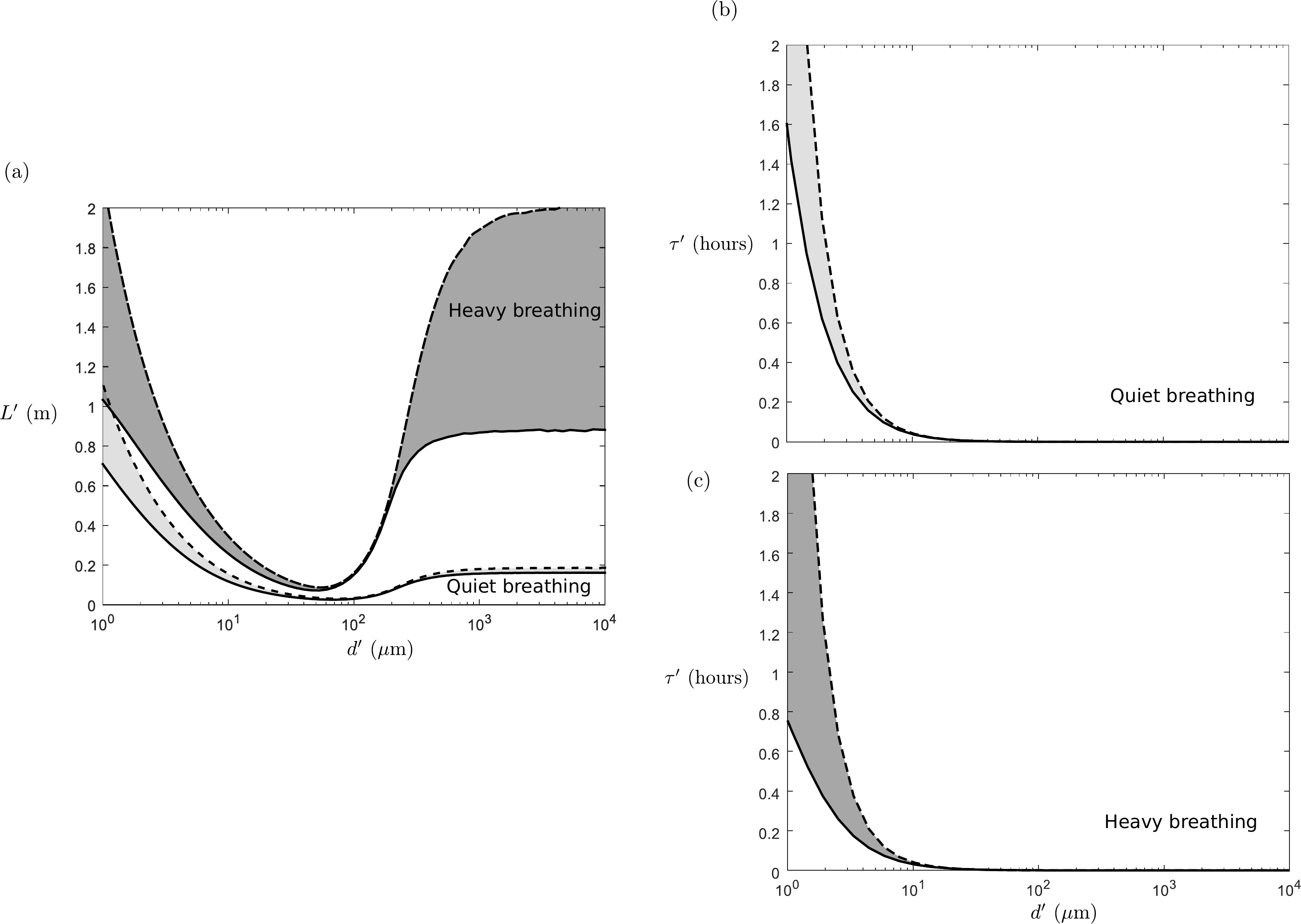}
\caption{(a) The maximum distance ($L'$)  travelled for droplets of various diameters ($d'$)  with quiet (light grey) and heavy (dark grey) breathing. (b,c) The total duration of travel $\tau'$ corresponding to the value of $L'$ shown in (a) for quiet breathing (b) and heavy breathing (c). The dashed curve corresponds to trajectories with with ejection angle equal to $\alpha+\beta/2$, while the solid curve corresponds to trajectories with ejection angle equal to $\alpha-\beta/2$ in Table~\ref{tab:dispersion}.  \label{fig:dispersion} }
\end{figure*}

We find that for both heavy and quiet breathing, the maximum distance travelled by droplets $L'$ (and the corresponding flight time $\tau'$) depends strongly on the droplet diameter -- see Figure~\ref{fig:dispersion}. As expected,  small droplets can travel many metres, however, we see that there is an intermediate range of droplet diameters where the horizontal distance is minimised. For quiet breathing, this minimum occurs between $69 \mu\text{m}<d< 77 \mu\text{m}$, while for heavy breathing this minimum occurs between $50 \mu\text{m}<d< 55 \mu\text{m}$. This multi-modal behaviour is reminiscent of that in previous experimental studies that measured the size distributions of droplets in various respiratory events such as talking  and coughing\cite{Xie2009,Johnson2011} and sneezing\cite{Han2013}. The multi-modal behaviour observed in experiments is attributed to the different generation modes: bronchiolar, laryngeal and oral. In our simplified model, we do not have any assumption on the biological origin of the droplet: the existence of the minimum is a characteristic of the \emph{droplets themselves} and cannot be used as an indicator of the underlying droplet size distribution. The time it takes $\tau'$ decreases monotonically with increasing droplet diameter as shown in Figure~\ref{fig:dispersion}b-c.

In order to unpick the physics, observe that the drag force scales with the diameter of the droplet, but the weight of the droplet scales with the diameter cubed, hence for large droplets, the drag force is negligible in comparison with the inertia of the droplet. As shown before, the droplets are slowed down exponentially in the horizontal direction and are accelerated in the vertical direction by gravity, giving the maximum horizontal range of the droplet (when nominally $Y=-1$)
\begin{equation}
X= (St/R)\left(1-\mathbf \exp{\left[-(R/St)\sqrt{\frac{2}{\left( 1-\frac 3 2 R\right) |g|}}\right]}\right).
\end{equation}
For large droplets ($St\gg R$) we can then estimate that the maximum distance $L=L'/|\mathbf x'_0|$ is
\begin{equation}
L\approx \sqrt{\frac{2}{\left( 1-\frac 3 2 R\right) |g|}},
\label{eq:max-large-St}
\end{equation}
meaning that the trajectories are ballistic, and we expect that for $St\gg R$, the maximum distance becomes independent of $St$, in agreement with the observation that large droplets' trajectories are independent of the surrounding flow\cite{Wei2015,Xie2007,Bourouiba2014}. 

For small droplets $St\ll R$, the drag decreases linearly with decreasing droplet diameter, but the weight rapidly decreases cubically with decreasing diameter. Hence small droplets follow the airflow faithfully with little influence from gravity. Such droplets can get great distances before falling, as shown in the left-hand side of Figure~\ref{fig:dispersion}.

In the case of small droplets, the horizontal component of the droplet's trajectory follows the airflow like a tracer, and the droplet falls at its Stokesian settling velocity. Upon inspection, we find that the maximum horizontal distance $L$ (when nominally $Y=-1$) tends to the following asymptote as $St\to 0$.
\begin{equation}
L=\left( \frac{2 A \Lambda}{\left(1-\frac{3}{2}R\right)|g|}\right)^{1/3}.    
\label{eq:max-small-St}
\end{equation}
We can therefore estimate that droplets for which $L>1$, or equivalently
\begin{equation}
St<\frac{2 R \Lambda}{\left(1-\frac{3}{2}R\right)|g|}, 
\label{eq:max-small-St-boundary}
\end{equation}
(i.e., the droplets travel farther in the horizontal direction than the vertical direction) weakly feel gravity. 

In between these two extreme cases, the drag force on the droplet is the same order of magnitude as the gravitational force. By balancing these two effects, we can approximate the upper bound of $St$ where the droplets become ballistic:
\begin{equation}
St< \frac{R}{\left( 1-\frac 3 2 R\right) |g| }.
\label{eq:ballistic}
\end{equation}
Such droplets are not light enough to get carried any great distance by the ambient airflow, but do not have large enough inertia to become ballistic.

\begin{figure*}
\includegraphics[width=\textwidth]{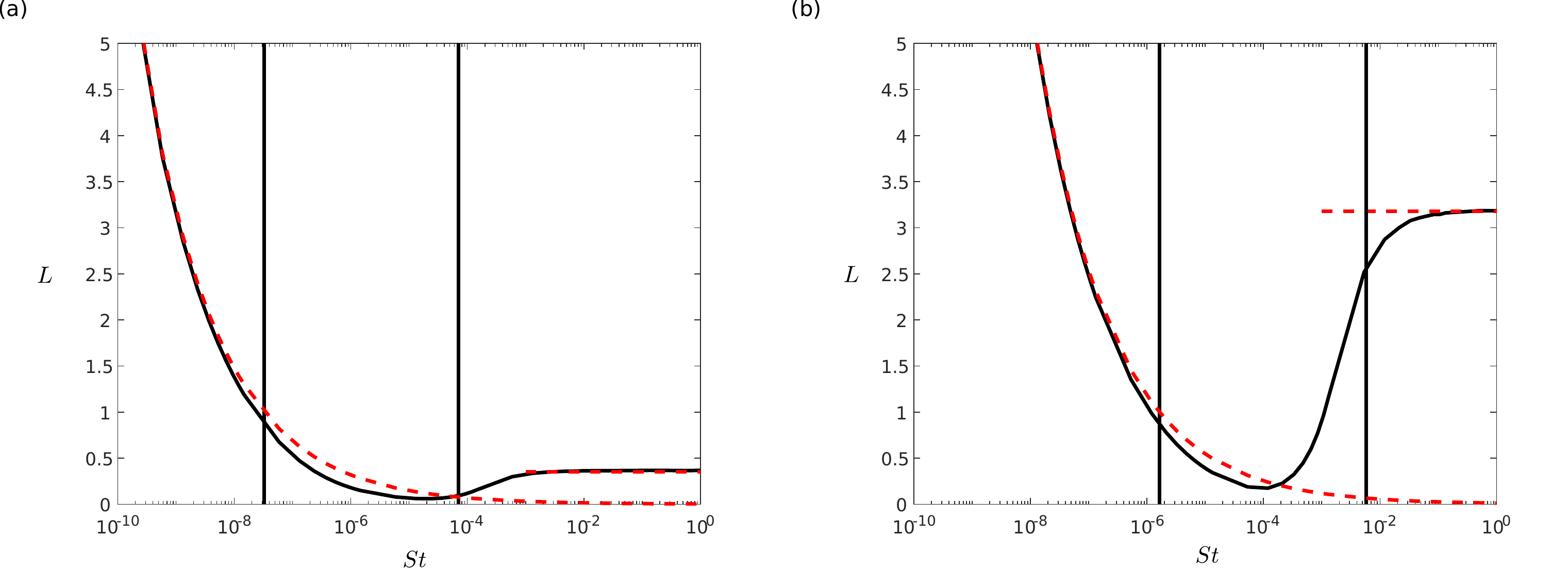}
\caption{The maximum distance ($L$) travelled for droplets of various $St$ with quiet breathing (a) and heavy breathing (b). The vertical solid lines indicate the distinction between small and intermediate-sized (from \eqref{eq:max-small-St-boundary}) and from intermediate-sized to large (from \eqref{eq:ballistic}). The red dashed curves indicate the small-$St$ \eqref{eq:max-small-St} and large-$St$ \eqref{eq:max-large-St} limits.
\label{fig:heavy-astmptotic} }
\end{figure*}

Hence, we have the following designations: 
\begin{enumerate}
    \item[(I)] small droplets with $St$ satisfying $St<\frac{2 R\Lambda}{\left(1-\frac{3}{2}R\right)|g|}$, which act like fluid tracers
    \item[(II)] intermediate-sized droplets with $\frac{2R\Lambda}{\left(1-\frac{3}{2}R\right)|g|}<St<\frac{R}{\left(1-\frac{3}{2}R\right)|g|}$.
    \item[(III)] large droplets with $St>\frac{R}{\left(1-\frac{3}{2}R\right)|g|}$, which adopt ballistic trajectories.
\end{enumerate} 
This is illustrated in Figure~\ref{fig:heavy-astmptotic}, where the black curves are the numerical solutions to quiet (a) and heavy (b) breathing at zero direction and spread angle\cite{Viola2020} and the red dashed curves indicate the expressions in \eqref{eq:max-large-St} for large $St$ and \eqref{eq:max-small-St} for small $St$. The black vertical lines indicate the distinction between small and intermediate-sized (see \eqref{eq:max-small-St-boundary}) and intermediate-sized and large droplets (see \eqref{eq:ballistic}). 

Reverting to dimensional quantities, we have the following range of intermediate-sized droplets.
\begin{equation}
 \sqrt{\frac{9 \nu'_{\text{air}} Q_1' \rho'_{\text{air}}}{ \pi  g' |\mathbf x_0'|^2( \rho'_{\text{drop}}- \rho'_{\text{air}})}}  <d'<2\sqrt{ \frac{9 \nu'_{\text{air}} \rho'_{\text{air}}  U'}{2g'( \rho'_{\text{drop}}-  \rho'_{\text{air}})}}.
\end{equation}
Plugging in the numbers from Table \ref{tab:dispersion}, we have the approximate range:
\begin{equation}
3\mu\text{m} <d'< 138 \mu\text{m},
\end{equation}
for quiet breathing and
\begin{equation}
7\mu\text{m} <d'< 414 \mu\text{m},
\end{equation}
for heavy breathing. Our upper bound is in good agreement with previous categorisations of droplets\cite{Chen2020}, although our lower bound seems to be smaller than those found by previous authors.

\subsection{The effectiveness of extraction on droplets}
\label{sec:extractor}

\begin{figure*}
\includegraphics[width=0.85\textwidth]{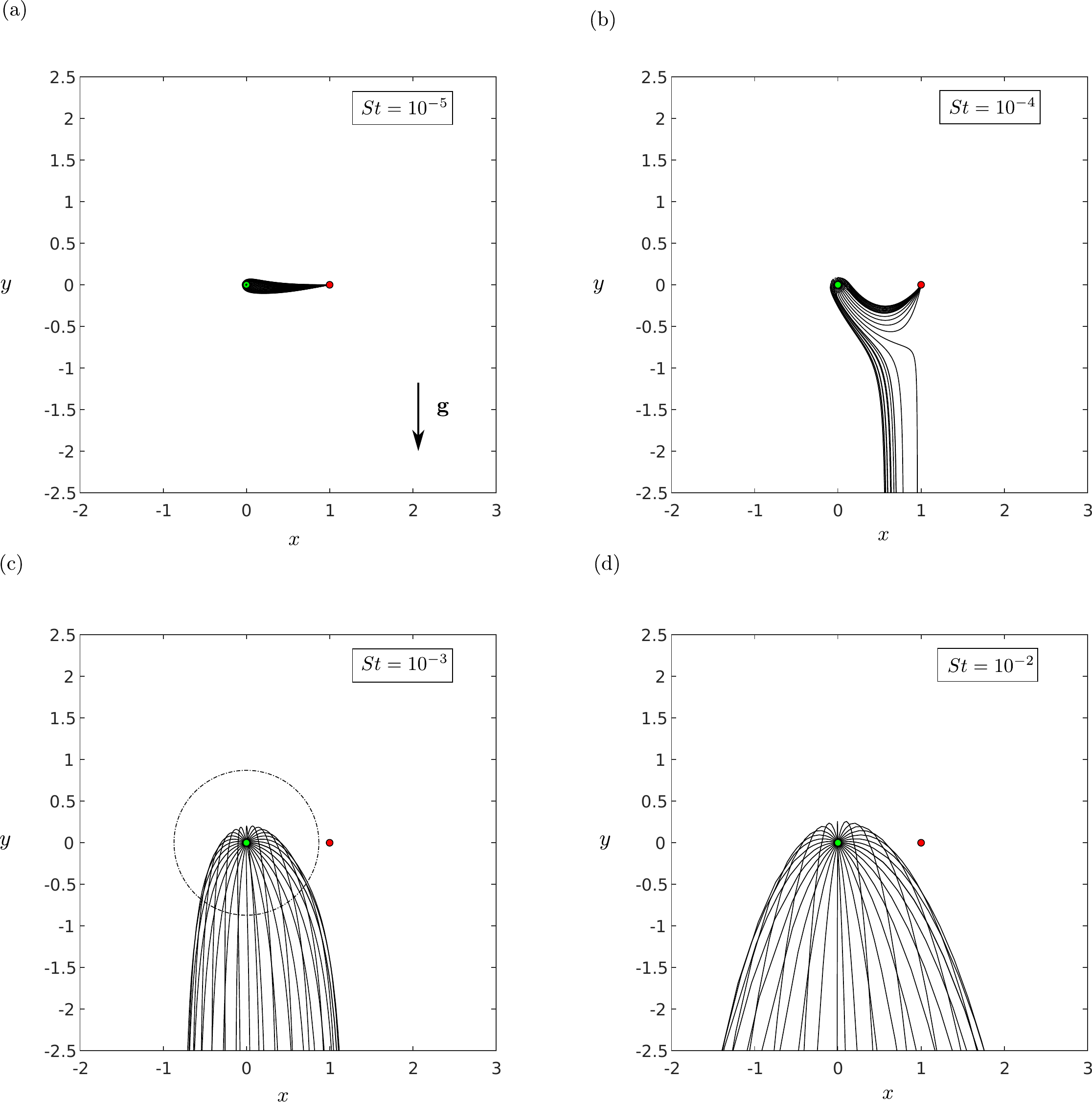}
\caption{\label{fig:gravity-study-St-extractor} The trajectories $\mathbf X(t)$ in the $xy$-plane with a background source-sink pair with strength ratio $\gamma=436$ and for various Stokes numbers $St$: (a) $St=10^{-5}$, (b) $St=10^{-4}$, (c) $St=10^{-3}$, and (d) $St=10^{-2}$. In these plots, $R=0.00115$, $\Lambda=0.00022$, and $|\mathbf g |=1.96$. The dash-dotted circle indicates the maximal distance predicted by inertia-drag balance. The source is indicated by a green filled circle and the sink is indicated by a red filled circle.}
\end{figure*}

Consider a person breathing air of density $\rho_{\text{air}}'=1.149\, \text{kg} \,\text{m}^{-3}$ and kinematic viscosity $\nu_{\text{air}}'=16.36\times 10^{-6}\, \text{m}^2 \,\text{s}^{-1}$ containing water droplets of density $\rho_{\text{drop}}'=1000\, \text{kg} \,\text{m}^{-3}$. In human respiration\cite{Asadi2019,Duguid1946}, the exhaled droplets have diameters $d'=2a'$ in the range 0.5 $\mu$m to 2000 $\mu$m. For a human breathing at rest, their average volume flux is in the range $Q_1'=$ 5-8 L min$^{-1}$: these values of flow rate are similar to those in previous studies\cite{Gupta2010}, which reports 13 litre/min for breathing, and the typical speed of a jet in normal breathing conditions is of the order of  $U'=$ 1 m s$^{-1}$. In violent respiratory events, such as sneezing or coughing, these values could be significantly higher\cite{Bourouiba2014}. Finally, the extraction unit is located a distance of $|\mathbf x_0'|=0.2$ m from the person. These quantities are summarised in Table \ref{tab:extractor}.

\begin{table}[]
    \centering
\begin{tabular}{llll}
\hline
Quantity    & Description & Value & Units \\
\hline
$U'$      & Breath jet velocity   & 1 &   m s$^{-1}$  \\
$\rho_{\text{air}}'$          & Density of air  & 1.149      & $\text{kg} \,\text{m}^{-3}$\\
$\rho_{\text{drop}}'$     & Density of droplet     & 1000     & $\text{kg} \,\text{m}^{-3}$ \\
$\nu_{\text{air}}'$      & Viscosity of air     & $16.36\times 10^{-6}$     & $\text{m}^2 \,\text{s}^{-1}$\\
$Q_1'$ & Volume influx      &6.5     & L min$^{-1}$ \\
$Q_2'$ & Volume outflux      &2832     & L min$^{-1}$ \\
$|\mathbf x_0'|$ & Characteristic length*      & 0.2     & m \\
\hline
\end{tabular}
\caption{Physical quantities for extraction. *The characteristic length is chosen to be the source-sink distance.} \label{tab:extractor}
\end{table}

Based on these numbers, the non-dimensional parameters that govern the trajectory of the droplet are determined to be $R=0.00115$, $Re=12,225$, $\Lambda=0.00022$, $|\mathbf g| = 1.96$, and the Stokes number ranges approximately from $10^{-9}$ to $10^{-1}$. The parameter $\gamma$ relates the flux of the extraction unit to the flux of a human's breath, and its effect will be examined. In particular, if we suppose that the envisaged extraction unit has a volume flux approximately equal to that of a standard vacuum cleaner ($2832$  L min$^{-1}$), then we can approximate that $\gamma\approx 436$. In Figure~\ref{fig:gravity-study-St-extractor}, we show the efficacy of such extraction for a range of $St$. Extraction is very effective at low $St$, however for $St>8.5\times10^{-5}$, such extraction is ineffective. This upper bound of the Stokes number corresponds to water droplets of diameter 71 $\mu$m. Droplets larger than this will not be collected by extraction. In the nomenclature of Section~\ref{sec:drop_disp}, the effective range of extraction corresponds to non-ballistic droplets.

\section{\label{sec:conclusion} Discussion and conclusion}

In this paper, we have presented a simplified mathematical model for droplet dispersion from a source and in the presence an aerosol extractor. In the absence of gravity, and for $St\ll R$, droplets behave as ideal tracers and the maximum distance that they can travel before being extracted is a function of $\gamma$ only. In this case, there are two (source, and sink if $\gamma=1$) or three (source, sink, and saddle if $\gamma>1$) fixed points. The fixed points in this study are co-linear, and the position of the saddle depends on $\gamma$ alone, for any given distance between source and sink. For moderate $St$, the droplets' inertia carry them far away from the source until they are slowed down by drag forces and pulled into the sink. In this case, the maximum distance that droplets can travel is given by $R/St$.

When gravity effects are taken into account, the saddle point for $St\ll R$ is no longer co-linear but moves on an arc, clockwise about the source, and a fourth fixed point (saddle) emerges approximately below the sink fixed point. For fixed $\gamma$, this fixed point moves closer to the source as the magnitude of gravity is increased. In this case, there is a set of trajectories that are pulled away from the sink by gravity. For moderate $St$, gravity plays an increasingly important role, and there is a critical value of gravity that pulls all trajectories vertically downwards away from the source. For yet larger $St$, the trajectories adopt a ballistic trajectory, with even those that travel close to the sink not being pulled in.

COVID-19 has brought increased awareness of the risks of aerosol generating procedures (AGPs) across all fields of medicine, highlighting the need for a deeper understanding of droplet dispersion and categorisation during respiration and AGPs. Clinicians recognise that our historical approaches to protection during AGPs are no longer adequate and that many additional precautions are necessary. In order to develop the most effective solutions, a critical first step is understanding the behaviour of droplets generated during AGPs. This paper, allows us to predict this behaviour and inform our understanding of “at risk” zones in the vicinity of an AGP. In particular, we performed simulations relevant to human respiration, as well as simulations to inform the development of an aerosol extractor for use in clinical settings. These simulations can help to guide recommendations on maximum safe distances between source and sink. 

Additionally, these models provide a better understanding of the behaviour of individual droplets of various sizes, that may be present in a wide range of aerosols contaminated with viruses or other pathogens. This may help clinicians to make better informed decisions regarding safety while performing AGPs; and in their choices of the type of PPE they wear. Lastly, these models provide a basis upon which aerosol and droplet contamination from a wide range of surgical, medical, dental and veterinary AGPs can be modelled, while taking into account airflows in confined clinical spaces.  In this case, we found that for $St \leq  8.5\times10^{-5}$, all of the aerosol is extracted and that gravity has minimal effect, this $St$ corresponds to droplets with approximate diameter equal to $0.07$ mm. Droplets larger than this are affected by gravity, and for $St=10^{-2}$, corresponding to droplets equal to $0.78$ mm, none of the droplets are extracted. Such large droplets would be typically captured by of personal protective equipment (PPE), such as FFP1 masks, which have pore sizes typically smaller than 1 $\mu$m.

We determined the maximum range of droplets ejected from the source in the absence of a sink, and found that the range is minimised for intermediate-sized droplets. We find that in human respiration, this pertains to droplets within the observed range of ejected droplets. This could have implications for the interpretation for data coming from experiments on biological subjects. In particular, those that attribute observed bi- and tri-modal droplet dispersion to biological functions. Our studies suggest that the bi-modal nature of the curve is a function of the droplet's Stokes number and not necessarily linked to a specific biological function.

In our model, we neglected the Basset history term in the Maxey-Riley equation. These Basset history term is of significant importance for bubbly flows, where it can account for a quarter of the instantaneous force on a bubble\cite{Michaelides1997}. Generally speaking, for $R\ll 2/3$, this term can be safely ignored for small and intermediate-sized droplets. Recent studies have also shown that neglecting it in modelling of raindrop growth leads to a substantial overestimate of the growth rate of the droplet. Hence, for the solutions that become ballistic, we expect that such trajectories would be influenced by the Basset history term, and should be included. To do this efficiently, there is a very promising method developed recently\cite{Prasath2019}. Since this is not the focus of our study (such droplets can be captured by other forms of PPE), we do not perform such a study here.

If the aerosol route of transmission is confirmed to be important by the World Health Organization\cite{WHO20,Busco2020}, we will need to reconsider guidelines on social distancing, ventilation systems, shared spaces, etc. To ensure that we put in place the correct mitigating measures such as, for example, face coverings, we need to have a better understanding of the different droplet behaviours and their different dispersion mechanisms depending on their size. This paper contributes to this debate by providing a new framework for categorising droplets depending on their dispersion mechanism.

\section*{\label{sec:acknowledgements} Acknowledgements}

OJA acknowledges the support of the Commonwealth Scholarship Commission.
FVM acknowledges the support of Caledonian Heritable Limited.

\section*{\label{sec:data-availability} Data availability}

The data that support the findings of this study are available from the corresponding author upon reasonable request.

\bibliography{aipsamp}

\end{document}